# Constraints for quasi-elastic neutron reflections in the range $\Delta E \sim 10^{-12} - 10^{-9}\,eV$.

V.V.Nesvizhevsky, Institut Laue-Langevin, Grenoble, France

## Abstract

A restrictive constraint for any quasi-elastic process was obtained in the previously inaccessible energy range $\Delta E \sim 10^{-12} - 3 \cdot 10^{-10}\,eV$ for reflections of ultracold neutrons from surfaces in the experiment on neutron quantum states in the earth's gravitational field. This could be useful for precision neutron spectrometry experiments and for the verification of extensions of quantum mechanics.

Corresponding author:
Valery Nesvizhevsky,
 Institut Laue-Langevin, 6 rue Jules Horowitz, F-38042, Grenoble, France
Tel.: 0476207795
Fax.: 0476207777
E-mail: nesvizhevsky@ill.fr

**Introduction**

One of the remarkable properties of ultracold neutrons (UCN) is their capacity for long storage in material traps with no significant change of their energy. This provides for a very broad range of applications for UCN in the investigation of neutron properties and their fundamental interactions: the measurement of the neutron lifetime [1-5]; the search for non-zero neutron electric dipole moment [6-7]; the study of the neutron quantum states in the gravitational field [8-9]; or the search for non-zero neutron electric charge [10] etc.

This precise conservation of UCN energy following their reflection from a surface is due to their long wavelength of $\sim 10^2$ A, which is two orders of magnitude larger than the average distance $\sim 1$ A between nuclei in reflecting matter. UCN are therefore reflected by the almost immovable average potential of a huge number of nuclei.

Residual thermal fluctuations of such a potential are extremely small due to the averaging of thermal fluctuations of a huge number of independent wall nuclei [11-12]. The probability of quasi-elastic scattering of UCN by the thermal fluctuations of a flat surface or by phonons in solids under normal conditions is expected to be equal to $10^{-14}$-$10^{-10}$ per collision. The energy change is $10^{-8}$-$10^{-7}$ eV - that is of the order of a typical effective potential. Such probability values are below present experimental sensitivity. The most sensitive upper limit was therefore measured at $\sim 10^{-8}$ [13] per collision. Quasi-elastic neutron reflections with an energy change of $10^{-12}$-$10^{-9}$ eV are much less probable than this.

More intensive quasi-elastic scattering was observed at more weakly-bound surface areas. Thus, for the recently discovered phenomenon of coherent quasi-elastic scattering of UCN off weakly-bound surface nanoparticles in a state of permanent thermal motion [13], the probability of such scattering was as high as $10^{-8}$-$10^{-2}$ per collision with a surface containing nanoparticles, while the average energy transfer was $10^{-8}$-$10^{-7}$ eV per collision, as expected.

In both the cases mentioned the energy of the main fraction of reflected neutrons did not change significantly. Moreover, for UCN storage at a flat solid surface free of nanoparticles, all reflections need to be highly elastic. This property of UCN reflection allows us to investigate any fundamental process which would change UCN energy.

In particular, as discussed in refs. [14-15], an extension of quantum mechanics with an additional logarithmic term in the Schrödinger equation assumes quasi-elastic scattering of UCN at the surface, with extremely small, but nevertheless measurable, energy changes. Such spectral measurements with UCN of high resolution were themselves methodologically challenging. They were also motivated by a long-standing anomaly in the storage of UCN in traps [16]. These experiments [17-18] allowed the authors to constrain such quasi-elasticity at $\sim 10^{-11}$ eV per collision, under the assumption of a "random walk" in phase space at each neutron collision with the wall: a non-zero result at this level was reported in ref. [17] at the limit of experimental sensitivity, but was not confirmed later in ref. [18], measured in the same setup with slightly better statistical sensitivity but with worse energy resolution.

A significant increase in the accuracy of neutron gravitational spectrometry was achieved in the recent experiment on quantum states of neutrons in the earth's gravitational field, using the high-resolution position-sensitive neutron detectors presented in ref. [19]. It has allowed us to improve many times over the upper limit for the probability and for the minimum energy transfer values for quasi-elastic scattering of UCN at the surface. Moreover, we can now consider energy changes at a single reflection, rather then having to follow the integral effects of many collisions, as in refs. [17-18]. In addition to this, the present limit concerns one specific component of the neutron velocity along the vertical axis before reflection and after it. Also any deviation from the conventional quantum mechanics can be verified in a more direct way in the quantum limit used here of the minimal possible initial energy, or velocity.

Such constraints, however, present a broader interest and could be considered in a more general model-independent way: how precisely do we know that UCN conserve their energy at wall reflections or during UCN storage in material traps?

**Experiment**

One should note that the present constrain was obtained as a by-product of the main experiment; the installation was not therefore optimized. It also contained extra components, not important for studying the quasi-elasticity (as, for instance, the mirror (2) in fig. 1). The experiment to study the quantum states of neutrons in the gravitational field, using position-sensitive neutron detectors [19-22], consists in the following.

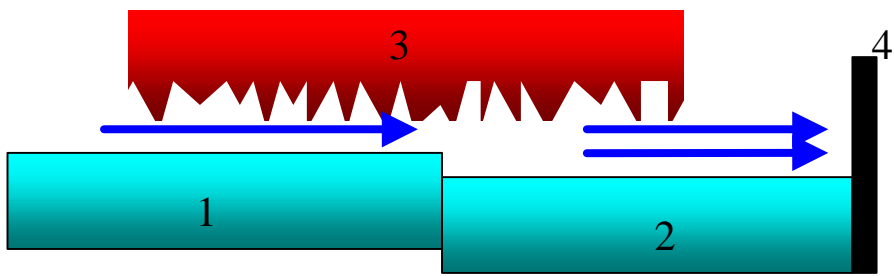

*Fig. 1. Scheme of the experiment to study neutron quantum states in the gravitational field using position-sensitive neutron detectors. 1,2 – mirrors; 3 – scatterer/absorber; 4 – position-sensitive neutron detector. Horizontal arrows illustrate the neutron trajectories. The mirrors are optical glasses polished to the roughness of ~10 A.*

A neutron beam with a horizontal velocity component of ~5 m/sec and a vertical velocity component of 1-2 cm/sec, which corresponds to the energy of the lowest neutron quantum state in the gravitational field above a mirror, is selected using a bottom mirror (1) and a scatterer/absorber (3) positioned above it at the height of ~20 μm. A second mirror (2) is installed lower by 21 μm than the first mirror (1). If the UCN bounce elastically on the mirror (2) surface in the zone between the scatterer's (3) exit edge and the position-sensitive detector (4), the measured spatial variation of the neutron density versus height would correspond to that shaped by the mirrors (1,2) and the scatterer (3) in the zone upstream of the scatterer's (3) exit edge. If they do not, then the excess number of neutrons observed in the higher position would be attributed to their quasi-elastic reflection from the mirror (2) surface. The experimental installation is designed in such a way that any known parasitic effects (vibration of the mirrors and the scatterer, residual magnetic field gradients, quasi-specular reflections of UCN from mirrors or at residual dust particles) should be small enough not to cause a significant change in the spectrum of vertical neutron velocities (see refs. [8-9,19-22]). The precision of the optical components' adjustment and the neutron detection resolution are equal to ~1 µm. The typical result of a few-days detector exposure in such an experiment is presented in fig. 2.

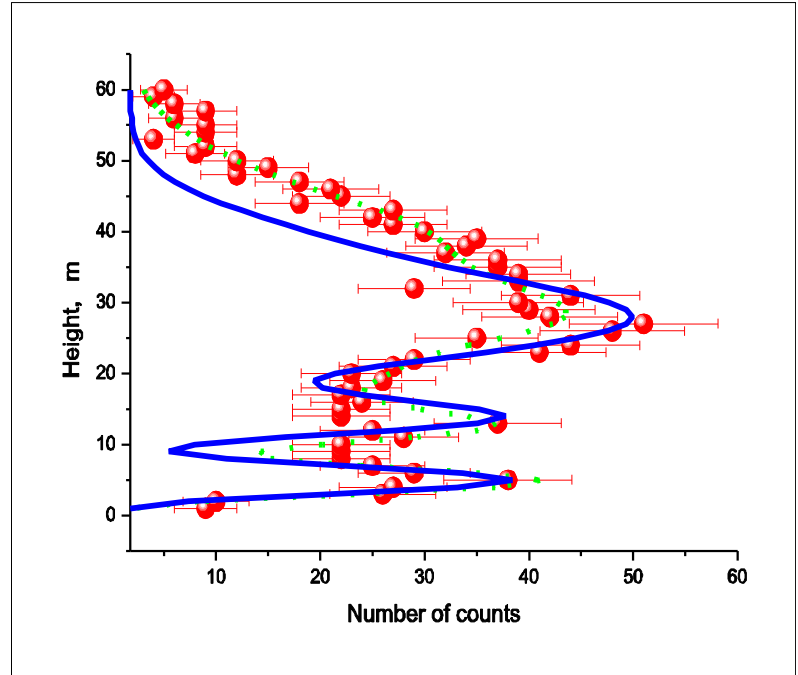

*Fig. 2. The neutron density distribution in the gravitational field is measured using position-sensitive detectors of extra-high spatial resolution. The circles indicate experimental results. The solid curve corresponds to the theoretical expectation under the assumption of an ideally efficient scatterer able to select a single quantum state above the mirror (1) and no parasitic transitions between the quantum states above the mirror (2). The dotted curve corresponds to the more realistic fit using precise wave-functions and free values for the quantum states populations. The detector background is constant in the range from -3 mm to +3 mm below and above the presented part of the detector.*

**Constraint on a quasi-elastic reflection**

We shall not discuss the physical reasons and conditions for possible quasi-elastic reflections of UCN at surfaces; we shall just consider this problem in phenomenological terms. A simple conservative upper limit for the quasi-elastic scattering/heating probability (versus average energy transfer) following UCN reflection from the lower polished glass mirror could be calculated, assuming an ideal scatterer able to select a single quantum state above the mirror (1) in fig. 1. Populations of all quantum states above the mirror (2) can be precisely calculated in this case [20]. They provide the neutron density distribution, presented by the solid curve in fig. 2. Actually we know that a few neutrons at higher quantum states should survive [22] producing the density distribution close to that presented by the dotted curve in fig. 2. However, we do not try to take such neutrons into account intentionally sacrificing the sensitivity of the present limit in favour of maximum reliability and transparency. Such an estimation could be further improved with the present experimental data using more sophisticated theoretical analysis based on ref. [22]. It would however be slightly model-dependent in such a case. For the simplified approach chosen, the solid line in fig. 2 is considered as "background" for the measurement of quasi-elasticity and any additional events above this line would be supposed to be due to quasi-elastic scattering. Fig. 3 illustrates the results of the treatment of the experimental data presented in fig. 2.

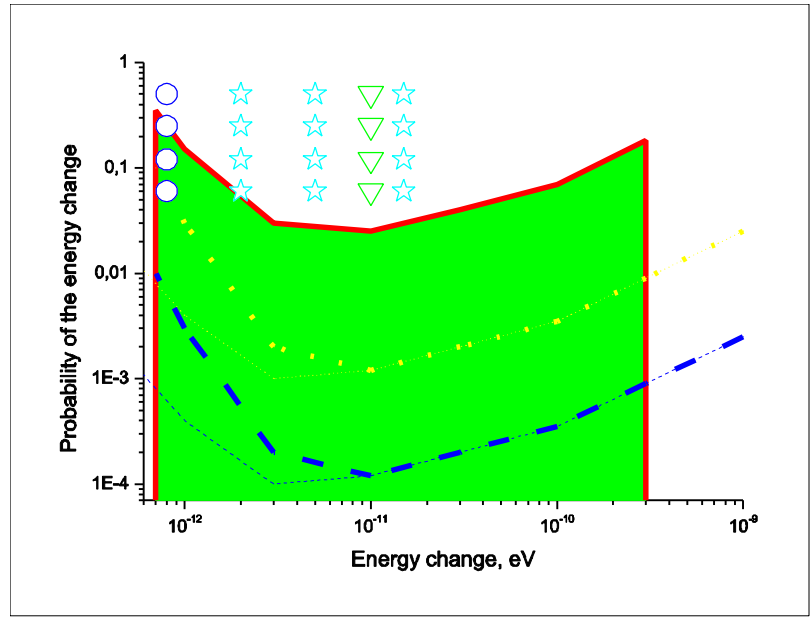

*Fig. 3. The solid curve corresponds to constraints for quasi-elastic scattering of UCN at a flat glass surface: the total probability of such a scattering per one quasi-classical bounce versus average energy transfer at "3s" confidence level. The dotted curve shows the possible improvement of such constraints in the flow-through measuring mode. The dashed curve indicates a further increase in sensitivity in the storage measuring mode. The circles correspond to theoretical predictions for the present experiment in accordance with refs. [14-15, 17]. The stars indicate analogous predictions for measurements with the experimental installation [8-9, 19-22] inclined to various angles. The triangles show the value of the energy change expected in refs. [14-15, 17] (for a higher initial neutron velocity than that in the present experiment). The thin dotted and dashed curves indicate schematically the constraints, for which the initial spectral shape line would be taken into account.*

The straightforward calculation of such a constraint provides the solid curve in fig. 3 under the following assumptions: 1) all additional events higher than the solid curve in fig. 2 are attributable to quasi-elastic scattering/heating; 2) the energy is assumed to change in one step (due to the low probability of such an event); 3) we take the number of quasi-classical collisions in such a system [22].

The rather sharp decrease with height of the neutron density at a characteristic scale of a few microns simplifies considerably the present calculation. For large enough $\Delta E$ values, any excess counts above the constant background level $\frac{\Delta N_{bg}}{\Delta h}$ in the height range $h > 60 mm$ are attributed to quasi-elastic scattering/heating. Quasi-elastically scattered neutrons could be observed at any height between zero and $\frac{E_0 + \Delta E}{m_n \cdot g}$, where $E_0$ is the initial energy of vertical motion, $\Delta E$ is the energy gain, $m_n$ is the neutron mass, and $g$ is the gravitational acceleration. If $\Delta E >> E_0$, the total number of background events is $\sim \frac{\Delta N_{bg}}{\Delta h} \cdot \frac{\Delta E}{m_n \cdot g}$, neglecting the initial spectral line width $h < 60 mm$. At "$3 \cdot s$" confidence level, we would observe an excess $N_{q.el.}$ of events at $h > 60 mm$, if it is equal to:

$$N_{q.el.} = 3 \cdot \sqrt{\frac{\Delta N_{bg}}{\Delta h} \cdot \frac{\Delta E}{m_n \cdot g}} \qquad (1)$$

With the horizontal velocity component $V_{hor}$ and the mirror length $L$ between the scatterer's exit edge and the detector (see fig. 1), the total number $N_{q.cl.}$ of quasi-classical bounces is:

$$N_{bounces} = \frac{L}{\frac{2}{g} \cdot \sqrt{\frac{2 \cdot E_0}{m_n}} \cdot V_{hor}} \qquad (2)$$

Thus, with the total number $N_0$ of neutrons in the initial spectral line, we would be able to observe quasi-elastic scattering at "$3 \cdot s$" confidence level if its probability $P_{q.el.}(\Delta E)$ is equal to:

$$P_{q.el.}(\Delta E) = \frac{N_{q.el.}}{N_0 \cdot N_{bounce}} = \frac{3}{N_0 \cdot L} \cdot \sqrt{\frac{\Delta N_{bg}}{\Delta h} \cdot \frac{\Delta E}{m_n \cdot g}} \cdot \frac{2}{g} \cdot \sqrt{\frac{2 \cdot E_0}{m_n}} \qquad (3)$$

As is evident from eq. (3), $P_{q.el.}(\Delta E)$ increases as $\sqrt{\Delta E}$, thus decreasing sensitivity of the present constraint at large energy changes. The sensitivity is also lower at energy changes smaller than the initial spectral line width of ~60 μm (here the constraint is estimated numerically). Therefore the best sensitivity is achieved at the energy change comparable to one or few initial spectral line widths, as shown in fig. 3.

**Analysis and prospects for strengthening constraint.**

The presented constraint shows the high degree of elasticity of neutron reflections in the range $\Delta E \sim 10^{-12}$-$3 \cdot 10^{-10}$ eV; this is important for the further development of precision neutron spectrometry experiments. Further improvements in the sensitivity of such constraints by an order of magnitude are feasible in the flow-through measuring mode, by improved shielding of the neutron detectors (a factor $\sqrt{\frac{\Delta N_b}{\Delta h}}$ in eq. 3), by increasing the length of the bottom mirror (a factor $\frac{1}{L}$ in eq. 3), by further increasing the scatterer efficiency, and by using of more narrow initial neutron spectrum (a factor $\sqrt{E_0}$ in eq. 3). On the other hand, a broader initial spectrum could allow us to increase a factor $N$ in eq. 3 and to improve therefore the sensitivity at higher $\Delta E$ values (sacrificing the sensitivity at lower $\Delta E$ values).

An almost order-of-magnitude gain in minimal measurable energy change could be achieved by providing a proper theoretical account (in accordance with ref. [22], for instance) of the spectrum-shaping properties of the scatterer, or by a differential measurement of the vertical spectrum evolution using bottom mirrors of different length. Possible improvements in the flow-through mode are illustrated by the dotted curve in fig. 3. One should note that any jumps in energy by the value significantly lower than 1 peV would clearly contradict to the observation of quantum states of neutrons in the gravitational field [8-9, 19-22] and therefore they are not analyzed in the present article. The minimal considered energy increase corresponds to the energy difference between neighboring quantum states in the gravitational field.

A much higher increase in sensitivity could be achieved in the storage measuring mode with the long storage of UCN at specular trajectories in a closed trap (the dashed curve in fig. 3 or better).

As an example of a possible application of the present constraint, let us compare it to the theoretical prediction in accordance with refs. [17-18]. This model assumes the replacement of "continuous interaction" of UCN with a gravitational field by a sequence of "collisions with the field". The time interval $dt$ between the "collisions" is defined as the time during which the mass "does not know that there is an interaction" since the kinetic energy change $dE$ (by falling) is too small to be resolved. From the uncertainty principle:

$$dt \cdot dE = \frac{\hbar}{2}, \text{ or } dE = \sqrt{\frac{\hbar \cdot m_n \cdot g \cdot V_{vert}}{2}} = \sqrt{33 \cdot V_{vert}} \, (peV) \qquad (4)$$

where $V_{vert}[m/s]$ and $peV = 10^{-12} eV$.

For the vertical velocity component $V_{vert} \sim 2.5 cm/s$ in our present experiment, the expected energy change is $dE \sim 8 \cdot 10^{-13} eV$ (shown as the circle in fig. 3). The "100%" probability of quasi-elastic scattering is slightly higher than the "$3 \cdot s$" experimental constraint (the solid line in fig. 3). However, considering the expected probability value of ~10% and low experimental sensitivity at small $\Delta E$ values, one needs to further improve the sensitivity of present constrain.

On the other hand, a slight modification of the experimental setup would allow one to verify clearly the considered hypothesis. Namely, the whole apparatus should be turned by a significant angle relative to the direction of the gravitational field. In this case, the vertical velocity component is comparable to the longitudinal velocity of 5-10 m/s. The transversal velocity component (relative to the bottom mirror) is very small, just equal to that in the experiment [8-9, 19-22]. All sensitivity estimations for quasi-elastic scattering/heating are analogous to those given above (see fig. 3). However, the theoretically predicted effect could be as high as $\sim 10^{-11}$ eV (as a function of the inclination angle) – just in the range of the best sensitivity of the present constraint: the stars in fig. 3. In order to measure a hypothetical cooling of UCN at their quasi-elastic reflections one should preliminary select a higher quantum state (n>1) and then to follow evolution of the corresponding neutron spectrum. The sensitivity estimations in the energy range $0 < \Delta E < E_0$ would be about as strong as those for the quasi-elastic heating if the experiment is optimized for this purpose. Such measurements would be significantly easier to perform than the measurement of the gravitationally bound quantum states because they do not require such record levels of energy and spatial resolution.

**Conclusion**

As a side-product of the experiment to study the neutron quantum states in the earth's gravitational field, we have constrained the quasi-elastic interaction of neutrons with flat solid surfaces in an energy transfer range of $10^{-12}$-$10^{-9}$ eV, a range not previously accessible for experiments with neutrons. This constraint could be useful for precision neutron spectrometry experiments, in particular for those using the neutron quantum states in the gravitational field [23], and also for the verification of various extensions of quantum mechanics. The constraint obtained from measurements with inclined bottom mirrors could reliably verify the hypothesis [14-15, 17] of the slight spectral evolution of UCN during their storage in traps. Further improvements in sensitivity by many orders of magnitude could be obtained, if they present further methodical or theoretical interest.

**Acknowledgement**

The author is grateful to prof. A.Steyerl for discussions and clarification of the theoretical hypothesis considered.